\documentclass{aa}

\pdfoutput=1 
\usepackage{graphicx}
\usepackage{latexsym,amsmath,amssymb}
\usepackage{hyperref}

\usepackage[authoryear]{natbib}
\bibpunct{(}{)}{;}{a}{}{,} 

\usepackage{txfonts}
\let\bibfont\undefined

\begin{document}


\title{The dark matter density at the Sun's location}

\author{Paolo Salucci\inst{1},
Fabrizio Nesti\inst{2},
Gianfranco Gentile\inst{3},
Christiane Frigerio Martins\inst{4}}

\institute{SISSA/ISAS, via Beirut 2-4, 34013 Trieste, Italy
\and
Dipartimento di Fisica,  Universit\`a di Ferrara, 44100 Ferrara, Italy
\and
Institut d'Astronomie et d'Astrophysique, Universit\'e Libre de Bruxelles, CP 226, Bvd du Triomphe, B-1050, Bruxelles, Belgium,\\ and 
  Sterrenkundig Observatorium, Universiteit Gent, Krijgslaan 281,
  B-9000 Gent, Belgium
\and
Universidade Federal do ABC, Rua Catequese 242, 09090-400
  Santo Andr\'{e}-S\~{a}o Paulo, Brasil}

\abstract                                                                                                                    
    {}                                                                                                                       
    {We derive the value of the dark matter density at the Sun's
      location ($\rho_\odot$) without globally mass-modeling the
      Galaxy.}
    {The proposed method relies on the local equation of centrifugal
      equilibrium and is independent of i) the shape of the dark
      matter density profile, ii) knowledge of the rotation curve
      from the galaxy center out to the virial radius, and iii) the
      uncertainties and the non-uniqueness of the bulge/disk/dark halo
      mass decomposition.}
    {The result can be obtained in analytic form and it explicitly
      includes the dependence on the relevant observational quantities
      and takes their uncertainties into account.  By adopting the
      reference, state-of-the-art values for these, we find
     $\rho_\odot=0.43(11)(10)\,$GeV/cm$^{3}$, where the quoted
      uncertainties are respectively due to the uncertainty in the
      slope of the circular-velocity at the Sun location and the
      ratio between this radius and the length scale of the stellar
      exponential thin disk.}
    {We obtained a reliable estimate of $\rho_\odot$, that, in
      addition has the merit of being ready to take into account any
      future change/improvement in the measures of the observational
      quantities it depends on.}

\authorrunning{P. Salucci et al.}                                                                                             
\titlerunning{Dark Matter density at the Sun's location}
\maketitle


\newcommand\sect[1]{\bigskip\noindent{\sl\bfseries #1.\ }}

\section{Introduction}

Galaxy rotation curves (e.g.~\citealt{rubin80,bosma81}) have unveiled
the presence of a dark ``mass component'' in spirals.  They are
pillars of the paradigm of massive dark halos, composed of a still
undetected kind of matter surrounding the luminous part of
galaxies. The kinematics of spirals shows universal
systematics~\citep{PSS,URC2}, which seems to be at variance with the
predictions emerging from simulations performed in the $\Lambda$ cold
dark matter ($\Lambda$CDM) scenario, (e.g.~\citealt{NFW96}), the
currently preferred cosmological paradigm of galaxy formation
(e.g.~\citealt{gentile04}).  Individual and coadded rotation curves
(RCs) of spiral galaxies are also crucial to investigate frameworks
alternative to the standard paradigm of cold collisionless DM in
Newtonian gravity (e.g.~\citealt{mond,bigr}).

At the same time, dedicated searches of DM particle candidates have
seen an important boost in recent years with relevant and costly
experiments being planned and executed.  The so-called
direct-detection experiments look for the scattering of DM particles
off the nuclei inside the detectors (e.g.\ CDMS, XENON10, DAMA/LIBRA)
by typically measuring the deposited energy or its annual modulation.
Clearly in all these experiments the signal is proportional to the DM
density in the Sun's region, $\rho_\odot$.  On the other hand
indirect-detection experiments (in particular Super-Kamionkande,
AMANDA, IceCube and ANTARES) search for the secondary particles
(neutrinos in these cases) produced by DM annihilations at the center
of the Sun or Earth, where it is expected that DM accumulates after
losing energy via scattering, possibly reaching a thermalized
state. The expected signal in this case depends on the DM density
inside these objects, which in turn is driven, via the capture
mechanism, by the same halo DM density in the Sun
region,~$\rho_\odot$.  Therefore, in both these kinds of direct and
indirect searches, an estimate of the the local density $\rho_\odot$ is
very important for a precise estimate of the signal or at least
reliable bounds on the DM cross-section vs mass to be compared with
limits from other searches.


What is then the value of $\rho_{\odot}$? A value of
\begin{equation}
\rho_\odot =0.3\, {\rm GeV/cm}^{3}
\label{eq:one}
\end{equation}
is routinely quoted in hundreds of papers, but how does this number
come out? In which works do we find the details of its measure?  It is
worth observing that in most of the cases in the literature, the above
value is given with no reference
(e.g.~\citealt{donato09,savage09}). Sometimes, the reference goes to a
couple of seminal papers. Among them, the Particle Data Group Review
\citep{pdg08} indicates the above value ``within a factor of two or
so'' and justifies it as coming from ``recent estimates based on a
detailed model of our Galaxy''.  However, the works cited are neither
recent nor detailed and sometimes not even an independent estimation
of $\rho_{\odot}$.

The only exception is the work by \cite{caldwell81} that devised what
can be considered as the standard method (CO hereafter) to determine
the value of $\rho_{\odot}$ from observations (see below). Their
resulting value, $0.23^{+0.23}_{-0.12}$GeV/cm$^{3}$, arises however
from very outdated kinematical observations and from a {\it cored}
(rather than a cuspy) halo distribution, so it is not a great support
for equation~(\ref{eq:one}). Similar conclusions can be drawn by
looking at other influential reviews: the papers they cite to back up
the value~(\ref{eq:one}) either do not estimate this quantity or use
very outdated observations.

\medskip
 
In general, it is quite simple to infer the distribution of dark
matter in spiral galaxies. Spiral's kinematics, in fact, reliably
traces the underlying gravitational potential \citep{PSS,URC2}. Then,
from coadded and/or individual RCs, we can build suitable mass models
that include stellar and gaseous disks along with a spherical bulge
and a dark halo. More in detail, by carefully analyzing (high quality)
circular velocity curves, with the help of relevant photometric and HI
data, one can derive the halo density at any desired radius.  The
accuracy of the ``measurements'' is excellent and the results are at
the core of the present debates on Galaxy formation
(e.g.~\citealp{gentile04,gentile05,deblok09}).

\medskip

To measure $\rho_\odot$, instead, is far from simple, because the MW
kinematics, unlike that of external galaxies, does not trace the
gravitational potential straightforwardly.  We do not directly measure
the circular velocity of stars and gas but rather, at our best, the
terminal velocity $V_T$ of the rotating HI disk, and this only inside
the solar circle (e.g.~\citealt{McClure07}). This velocity is related
to the circular velocity $V(r)$, for $r<R_{\odot} $, by means of
$V(r)= V_T(r)+ V_{\odot} \ r/R_{\odot}$, where $R_\odot\simeq 8$ kpc
is the distance of the Sun from the Galaxy center and $V_\odot$ the
value of the circular velocity at the Sun's position.  Both quantities
are known within an uncertainty of 5\% - 10\% (e.g.~\citealt{MB09}),
which triggers a similar uncertainty in the derived magnitude and
slope of the circular velocity.

As a result, and also considering other kinematical observations
(see~\citealt{sofue1}), the circular velocity of the MW from 2 to 8 kpc
can be only derived within non negligible uncertainties:
 \begin{eqnarray}
\label{eq:Vcirc}
&&V(r)=(215 \pm 30) \ {\rm km/s}\\
&&d{\rm log} V(r)/d{\rm log}r  \equiv \alpha(r) = 0.0 \pm  0.15\,.
\label{eq:slope}
\end{eqnarray}
Note that the range of circular velocities in equation
(\ref{eq:Vcirc}) is created by a mix of  a)  observational errors,
b) uncertainties in the values of $R_\odot$ and $V_\odot$, and c) actual 
radial variations of  $V$.  The first two trigger also part of the
range of the velocity slope (\ref{eq:slope}).  We stress that data
show that the radial variations of $\alpha(r)$ are small and likely
caused by the uncertainties just discussed: in either case we have
$d\alpha(r)/dr\simeq0\pm 0.03/$kpc $\simeq 0$.  In other words, in
this region the RC can be approximated by a straight line, whose slope
is known however only within a degree of uncertainty.

The outer (out to 60 kpc) MW ``effective'' circular velocity
$V(r)=(GM(r)/r)^{1/2}$ is, instead, much more uncertain and depends
on the assumptions made on dynamical and structural properties of its
estimators.  It appears to decline with radius, with quite an uncertain
slope~\citep{battaglia05,xue08,brown09}
\begin{equation}
  d{\rm log} V(r)/d{\rm log} r  = -0.20^{+0.05}_{-0.15}\qquad(R_\odot<r<60\,\text{kpc})\,.
\end{equation}
These uncertainties, combined with the intrinsic ``flatness'' of the
RC in the region specified above (that complicates the mass modeling
even in the case of a high-quality RC~\citep{Tonini}), make it very
difficult to obtain a reliable bulge/disk/halo mass model and
consequently an accurate estimate of~$\rho_\odot$.
 
To overcome these serious difficulties Caldwell and Ostriker
(\citeyear{caldwell81}) developed a method in which other
observational data, linked in various ways to the gravitational
potential, help with the mass modeling. These include the l.o.s.\
dispersion velocities of bright tracers at known distances from the
Sun (e.g.\ OB stars) and the total Galaxy mass. This extra information
allows a determination of $\rho_\odot$, though it turns out to be
uncertain within a factor~2 \citep{caldwell81}, or somewhat less when
more constraints from the $z$ motions of disk stars are added
(e.g.~\citealt{Olling,weber09,sofue1}).\footnote{To overcome the
  observational limits of equation\ (\ref{eq:slope})\, by resorting to
  additional tracers of the MW gravitational potential also requires
  taking extra assumptions, and the complex procedure can trigger some
  mild internal inconsistency in the resulting MW mass model (see
  Appendix~\ref{app:CU}).}

By averaging the results from different determinations obtained so
far, one finds $\rho_\odot=(0.3 \pm 0.2)\, {\rm GeV/cm}^{3}$, where
the uncertainty in the result is triggered in an unspecified way from
observational and fitting uncertainties, as from biases arising from
the various assumptions taken by the method, not the least the assumed
DM density profile.

Recently, by applying the CO method with a refined statistical
analysis to a large set of observational data, \cite{cu09} claimed a
measure with a very small uncertainty: $\rho_\odot=(0.389 \ \pm
0.025)\,$GeV/cm$^{3}$. While this result would be noticeable, it has
not been confirmed by a subsequent work~\citep{weber09} and it seems
unlikely, in view of equations (\ref{eq:Vcirc}) and (\ref{eq:slope}),
reflecting the state of art of our (lack of) knowledge.

\medskip

The aim of this work is to derive $\rho_\odot$ by following a more
direct route than mass modeling the whole Galaxy and dynamically
modeling a number of galactic components and a series of galactic
potential tracers. This will be done by means of a specifically devised
method and by using some recent results obtained for external
galaxies \citep{URC2}.  The idea is to resort to the equation of
centrifugal equilibrium, holding in spiral galaxies (see~\citealt{fall80}
for details)
\begin{equation}
V^2/r = a_H+a_D+a_B\,,
\label{eq:vtot}
\end{equation}
where $a_H$, $a_D$, and $a_B$ are the radial accelerations generated
by the halo, stellar disk, and bulge mass distributions.  Taking first
the (quite good) approximation of spherical DM halo, we have $a_H
\propto r^{-2}\int_0^r \rho_H(R) \ R^2 dR$. A similar relation holds
for the bulge.  Therefore, by differentiating
equation~(\ref{eq:vtot}), we obtain the DM density at any radius in
terms of the local angular velocity $\omega(r)=V/r $, the RC slope
$\alpha(r)$, the disk-to-dynamical mass ratio $\beta(r)$ (see later),
and the bulge mass density:
\begin{equation}
\rho_H (r) =   \omega(r)^2 [F_{tot}(\alpha(r)) - F_D (\beta(r)]-\rho_B(r).
\label{eq:rhoDM1}
\end{equation}
with $F_{tot}$ and $F_D$ known functions. 

In spirals, equation~(\ref{eq:rhoDM1}) is not useful for determining
the DM density at any radius because 1) it virtually collapses for
$r<R_D$ where $F_{tot} \simeq F_D$, and the bulge density can also
become dominating, $\rho_B\gg \rho_H$; 2) the radial variations of
$\alpha(r)$ have non-negligible observational uncertainty that further
complicates the effect discussed in the previous point; 3) the
quantity $\omega$ is known with less accuracy than $V$, the
observational quantity entering the traditional mass modeling.

Instead, in estimating $\rho_\odot$, i.e.\ the density of the MW DM
halo at a {\it specific} radius (the Sun position), the above
drawbacks disappear: 1) since $R_\odot>3 R_D$ we have
$F_{tot}(R_\odot)\gg F_D(R_\odot)$, equation~(\ref{eq:rhoDM1}) does
not collapse and, as a bonus, the most uncertain term of the r.h.s.\
of (\ref{eq:rhoDM1}) is also the smaller one; 2) $\omega_\odot$ is
very precisely measured; 3) $d \alpha/dr|_{R_\odot}\simeq 0$; and 4)
at the Sun's position the bulge density $\rho_B(R_\odot)$ is totally
negligible, $< \rho_H/50$ (e.g.~\citealt{sofue2}).  Thus, this method
is very powerful for determining the value of the DM density at
$R_{\odot}$.  The DM density at {\it any} radii is obviously left to
the standard mass modeling.

The method is obviously simpler for a spherically symmetric DM halo,
and can be further simplified by considering an infinitesimally thin
disk for the distribution of stars in the Galaxy.  However, below we
also include the effects of a possible halo oblateness and disk
thickness. Here, we anticipate that these effects, constrained by
observations, are rather weak, of the order of a few percent, and
are therefore irrelevant for this work.  As a result, we obtain a
reliable and model-independent determination of the local DM halo
density, as well as of its intrinsic uncertainty.

\medskip

In the next section we describe the method in detail and derive
$\rho_\odot$ and its uncertainty as a function of the relevant
observables.  In the last section we discuss the results and draw the
conclusions. In the Appendices, we explicitly describe the effect of
the halo oblateness and disk thickness, and we comment on the inherent
problems in the traditional determination of $\rho_\odot$.

\section{A model-independent method}

We model the Galaxy as composed by a stellar exponential thin disk
\citep{Freeman}, plus an unspecified spherical DM halo with density
profile $\rho_H(r)$. For the present work we can neglect the HI disk
because its surface density, between $2\,$kpc and $R_\odot$, is 100 to
5 times smaller than the stellar surface
density~\citep{nakanishisofue}.  Similarly, we neglect the stellar
bulge because, as mentioned above, its spatial density at $R_\odot$ is
virtually zero (e.g.~\citealt{sofue2}). The standard method and its
variants cannot take these very simplifying assumptions because the
global modeling of the Galaxy involves these mass components in a
crucial way.

As discussed in the introduction we can rewrite the equation of
centrifugal equilibrium by subtracting the disk component from the
total acceleration. From its radial derivative we then find
\begin{equation}
\rho_H(r)=\frac{X_q}{4 \pi G r^2}\, \frac{d}{dr}\left[r^2\left(\frac{V^2(r)}{r}-a_{\rm D}(r)\right)\right],
\label{eq:rhohalo}
\end{equation}
where $X_q$ is a factor correcting the spherical Gauss law used above
in case of oblateness $q$ of the DM halo. 

We describe the factor $X_q$ in appendix~\ref{sec:obladi}.  Since the
observational evidence for an oblateness of the Galactic DM halo is
still quite uncertain and at the same time the mean value of $q$, as
measured in a very large sample of external spirals, is very near to 1
(see e.g.~\cite{OBrien:2010eg} for a recent review of this issue), we
consider $q=0.95 \pm 0.05$ here as the reference value and compute
$X_q$ accordingly (see Appendix~\ref{sec:obladi}).  We thus find that
from present observations $X_q$ boils down to a correction of less
than 5\% to $\rho_\odot$, $X_q\simeq1.00$--$1.05$. Let us stress that
by equation (\ref{eq:obl})\ one may also take into account a value of
$q$, emerging from future improved observations, outside the range
considered here.

The disk component can be reliably modeled as a Freeman stellar
exponential thin disk of length
scale~\citep{PR04,juric08,robin08,reyle09} $R_{\rm D}=(2.5\pm 0.2) \
{\rm kpc}$. The stellar surface density is then: $\Sigma(r) = (M_{\rm
  D}/2\pi R_{\rm D}^2)\,e^{-r/R_{\rm D}}$.  Also, the disk can be
considered infinitesimally thin. In fact, its thickness $z_0$ is
small, $z_0 \sim 250\,$pc \citep{Juric:2005zr} and moreover $z_0\ll
R_D<R_\odot$, so that its effect on the derivative of the
acceleration, and in turn on our measure, is very limited.  For the
sake of completeness, we compute it explicitly in
Appendix~\ref{sec:zorro} and show that it implies a reduction of less
than 5\% of the $a_D$ term as computed for an infinitesimally thin
disk.  We can thus write $a_{\rm D}(r)= \frac{GM_{\rm D}r}{R_{\rm
    D}3}(I_0 K_0 - I_1 K_1)\, X_{z_0}$, where $I_n$ and $K_n$ are the
modified Bessel functions computed at $r/2R_{\rm D}$, and
$X_{z_0}\simeq0.95$ accounts for the nonzero disk thickness (See
Appendix~\ref{sec:zorro}).

Since only the first derivative of the circular velocity $V(r)$ enters
in (\ref{eq:rhohalo}) and in any case this function in the solar
neighborhood is almost linear, we can write
\begin{equation}
V(r)=V_{\odot} [1+ \alpha_\odot\, (r-R_\odot)/R_\odot ]\,,
\end{equation}
where $\alpha_\odot=\alpha(R_\odot)$ is the velocity slope at the Sun's
radius. Then equation (\ref{eq:rhohalo}) becomes
\begin{equation}
\rho_H(r)= \frac{X_q}{4 \pi G}\left[\frac{V^2(r)}{r^2} (1+2 {\alpha_\odot})
  -\frac{GM_{\rm D}}{R_{\rm D}^3} H(r/R_{\rm D})\,X_{z_0}\right],
\label{rhoh}
\end{equation}
with $ 2H(r/R_{\rm D})= (3I_0 K_0 - I_1 K_1) + (r/R_ D)(I_1 K_0 - I_0
K_1)$.  Equation~(\ref{rhoh}) holds at any radius outside the bulge
region and measures $\rho_H( R_\odot) \equiv \rho_\odot$ by
subtracting the ``effective'' density of the stellar component from
the one of all the gravitating matter.
 
The disk mass can be parametrized~\citep{PS90} by  $M_{\rm D}=
\beta \ 1.1 \ G^{-1} V^2_\odot R_\odot$, with $\beta=V_{\rm
  D}^2/V^2|_{R_\odot} $, i.e.\ the fraction of the disc contribution
to the circular velocity at the Sun.

Finally, by exploiting the fact that the quantity
$V/R|_{R_\odot}\equiv \omega = (30.3 \pm 0.3) \ {\rm km/s/kpc}$ is
measured with very high accuracy and much better than $V_\odot$ and
$R_\odot$ separately \citep{MB09,reid09}, and after defining $r_{\odot
  D}\equiv R_\odot/R_{\rm D}$, we obtain
\begin{equation} 
\rho_\odot =  1.2 \times 10^{-27} {\rm
 \frac{ g}{cm^3}}\left(\frac{\omega}{{\rm km/s\,kpc}}\right)^2 \!X_q\bigg[ (1 + 2
{\alpha_\odot}) - 1.1 \,\beta \, f(r_{\odot D})\, X_{z_0} \bigg]\,,
\label{rhoodot}
\end{equation}
where $f(r_{\odot D})=r_{\odot D}^3H(r_{\odot D})$.

It is now possible to observe the advantages of the proposed method:
a) it does not require assuming a particular DM halo density profile,
or the dynamical status of some distant tracers of the gravitational
field; b) it is independent the (poorly known) values of $V_\odot$ and
of the RC slope at different radii; c) it does not depend on the
structural properties of the bulge, which in the mass modeling creates
a degeneration with the stellar disk and DM halo. d) it only mildly
depends on the ratio $r_{\odot\rm D}$, as well as on the disk mass
parameter $\beta$; finally, e) the method depends on the RC slope at
the Sun ${\alpha_\odot}$, although in a specified way.  In all points
a) -- e) the method brings an evident improvement over the CO one.

\medskip

To proceed further we discuss the parameters appearing in
equation~(\ref{rhoodot}). Our determination does not depend on the
value of $R_\odot$ and $R_D$ separately, but only on their ratio
$r_{\odot D}$. For this we adopt the reference value and uncertainty
$r_{\odot D}= 3.4 \pm 0.5$,\footnote{We anticipate that
    $\rho_\odot$ is determined here for {\it any } reasonable value of
    $r_{\odot D}$ independently of the values taken today for $R_D$
    and $R_\odot$.} as suggested by the values of $R_D$ given above
and by the average of values of $R_\odot$ obtained in recent work
$R_\odot=8.2\pm0.5\,$kpc (\cite{ghez,Gillessen:2008qv,bovy}).  This
leads to $f(r_{\odot D})\simeq 0.42\pm 0.20$, whose uncertainty
propagates only mildly into the determination of $\rho_\odot$, because
the second term of the r.h.s.\ of equation~(\ref{rhoodot})\ is only
one third of the value of the first. (And it can only reach one half
by stretching all other uncertainties.)

Present data constrain the slope of the circular velocity at the Sun
to a central value of ${\alpha_\odot}=0$ and within a fairly narrow
range $ -0.075 \leq {\alpha_\odot} \leq 0.075$.  The uncertainty of
$\alpha_\odot$ is the main source of the uncertainty of the present
determination of $\rho_\odot$, and let us recall that also values
outside our adopted range may be used in the present analytic
determination, for instance, a value belonging to a wider range allowed
for ${\alpha_\odot}$ claimed by \citep{Olling}.

In equation~(\ref{rhoodot}), $\beta$ is the only quantity that is not
observed and therefore intrinsically uncertain. We can, however,
constrain it by computing the maximum value $\beta^M$ for which the
disk contribution at $2.2 \,R_D$ (where it has its maximum) totally
accounts for the circular velocity.  With no assumption on the halo
density profile one gets $ \beta^M=0.85$, independently of $V_\odot$
and $R_\odot$~\citep{PS90}.  However, this is really an absolute
maximal value and it corresponds, out to $R_\odot$, to a solid body
halo profile: $V_{h}\propto R^{\alpha_h}$ with $\alpha_h=1$.  Instead,
all mass modeling performed so far for the MW and for external
galaxies have found a lower value $ \alpha_h (3 R_{\rm D})\leq 0.8$,
which yields $\beta^M=0.77$. We can also set a lower limit for the
disk mass, i.e.\ $\beta^m$: first, the microlensing optical depth to
Baade's Window constrains the baryonic matter within the solar circle
to be greater than $3.9\, 10^{10} M_\odot$~\citep{MB09}.  Moreover,
the MW disk B-band luminosity $L_{\rm B} =2 \times 10 ^{10} L_{\odot}
$ coupled with the very reasonable value $M_D/L_B =2 $ again implies
$M_D \simeq 4 \, 10^{10} M_\odot$. All this implies $ \beta^m=
\beta^M/1.3\simeq 0.65$.\footnote{While these constraints of the disk
  mass reduce the uncertainty in the present determination of
  $\rho_{\odot}$, they improve the performance of the traditional
  method very little, where the uncertainties in the disk mass value
  do not trigger the most serious uncertainties of the mass modeling,
  as discussed in the Introduction.} We thus take
$\beta=0.72^{+0.05}_{-0.07}$ as reference range.

\smallskip

Using the reference  values, we get 
\begin{eqnarray}
 \rho_{\odot}\!&=&\!0.43 {\rm \frac{GeV}{cm^3}}\Bigg[1
+ 2.9\,{\alpha_\odot}
-0.64\, \bigg(\beta-0.72\bigg)
+0.45\bigg(r_{\odot D}-3.4\bigg) \nonumber\\
&&{}\qquad\qquad 
- 0.1\left(\frac{z_0}{\text{kpc}}-0.25\right)
+0.10\,\bigg(q-0.95\bigg)\nonumber\\
&&{}\qquad\qquad
+0.07\left(\frac{\omega}{\rm km/s\,kpc}-30.3\right)\Bigg]\,.
\label{eq:10bis}
\end{eqnarray}
This equation, which is the main result of our paper, estimates the DM
density at the Sun's location in an analytic way, in terms of the
involved observational quantities at their present status of
knowledge.  The equation is written in a form such that, for the
present reference values of these quantities, the term in the square
brackets on the r.h.s equals 1, so that the central result is
$\rho_\odot=0.43\,$GeV/cm$^3$. As such, the determination is ready to
account for future changes, improved measurement or any choice of
${\alpha_\odot}$, $\beta$, $z_0$, $\omega$, $r_{\odot D}$, $q$
different from the reference values adopted here, by simply inserting
them in the r.h.s.\ of eq (\ref{eq:10bis}).

\medskip

The next step is to estimate the uncertainty in the present
determination of $\rho_\odot$, which is triggered entirely by the
uncertainties of the quantities entering the determination.  From
equation (\ref{eq:10bis}) and the allowed range of values discussed
above, we see that the main sources of uncertainty are
${\alpha_\odot}$, $\beta$ and $r_{\odot D}$, which appear in the first
line. The other parameters give at most variations of 2-3\%, and can
be neglected in the following.

Then, first, it is  illustrative to consider ${\alpha_\odot}$,  $\beta$ and
$r_{\odot D}$ as independent quantities.  We thus have:
\begin{equation}
 \rho_{\odot}=\bigg(0.43 
\pm 0.094_{({\alpha_\odot})}
\mp 0.016_{(\beta)}
\pm 0.096_{(r_{\odot D})}
\bigg) {\rm \frac{GeV}{cm^3}}
\,,
\label{eq:result2}
\end{equation}
where $A_{(x)}$ means that $A$ is the total effect due to the possible
span of the quantity $x$.

\medskip

At this point, we can go one step further, assuming that the MW is a
typical spiral, and using recent results for the distribution of
matter in external galaxies, namely that DM halos around spirals are
self similar~\citep{URC2} and that the fractional amount of stellar
matter $\beta$ shapes the rotation curve slope
${\alpha_\odot}$~\citep{PS90b}:
\begin{equation}
\beta= 0.72 - 0.95\, {\alpha_\odot}\,.
\end{equation}
Using this relation in equation (\ref{eq:10bis}) we find (neglecting the
irrelevant $q$ and $z_0$ terms)
\begin{eqnarray}
 \rho_{\odot}\!&=&\!0.43 {\rm \frac{GeV}{cm^3}}\Bigg[1
+ 3.5\,{\alpha_\odot}
+0.45\bigg(r_{\odot D}-3.4\bigg)+
 \nonumber\\
&&{}\qquad\qquad\qquad\qquad
+0.07\left(\frac{\omega}{\rm km/s\,kpc}-30.3\right)\Bigg]\,.
\label{eq:11bis}
\end{eqnarray}
From the current known uncertainties, with the estimated range of
${\alpha_\odot}$, we find
\begin{equation}
 \rho_{\odot}=\bigg(0.430 
\pm 0.113_{({\alpha_\odot})}
\pm 0.096_{(r_{\odot D})}
\bigg) {\rm \frac{GeV}{cm^3}}\,.
\label{eq:result3}
\end{equation} 
This is our final estimate, which is somewhat higher than previous
determinations. Its uncertainty mainly reflects our poor knowledge of
the velocity slope ${\alpha_\odot}$ and the uncertainty in the galactocentric
Sun distance.

\section{Discussion and conclusion}
In this work we have provided a model-independent kinematical
determination of $\rho_\odot$.  The method proposed here derives
$\rho_\odot$ directly from the solution of the equation of centrifugal
equilibrium, by estimating the difference between the `total' density
and that of the stellar component.

The method leads to an optimal kinematical determination of
$\rho_\odot$, avoiding model-dependent and dubious tasks mandatory
with the standard method, i.e., a) to assume a particular DM density
profile and a specific dynamical status for the tracers of the
gravitational potential, b) to deal with the non-negligible
uncertainties of the global MW kinematics, c) to uniquely disentangle
the flattish RC into the different bulge/disk/halo components.

While the measure of $\rho_\odot$ can be performed in an ingenious
way, it cannot escape the fact that it ultimately depends at least on
three local quantities, the slope of the circular velocity at the Sun,
the fraction of its amplitude due to the DM, and the ratio between the
Sun galactocentric distance and the disk scale-length, whose
uncertainty unavoidably propagates in the result.

Two of these three quantities can be related by noting that the MW is
a typical Spiral and using the relations available for these kind of
galaxies~\citep{URC2}, so that the final uncertainty can be slightly
reduced.

We found that some oblateness of the DM halo and the small finite
thickness of the stellar disk play a limited role in the measure.
However, we took them into account by the simple correction
terms described.

The resulting local DM density that we find, $\rho_{\odot}=(0.43 \pm
0.11_{({\alpha_\odot})} \pm0.10_{(r_{\odot D})})\, {\rm GeV/cm^{3}}$,
is still consistent with previous determinations, or slightly higher.
However, the determination is free from theoretical assumptions and
can be easily updated by means of equation~(\ref{eq:10bis}) as the
relevant quantities will become better known.\footnote{Again, in the
  traditional method most of the uncertainty in the measure of
  $\rho_\odot$ discussed in the Introduction cannot be overcome by
  having more and better data.}

A final comment is in order. The {\it values} of $\rho_\odot$ found in
previous studies by means of the traditional methods (e.g.\
\citealt{sofue2, weber09}) differ among themselves and also from the
present value only by a small factor.  This relatively good agreement
in the values does not imply a concordance in the underlying mass
models, in the various assumptions taken or in the data set
employed, but is mainly due to the fact that $\omega$ or equivalently
$A-B$ (the well known combination of Oort constants) is measured with
good precision.  In fact, from $M_H(r)\propto V^2_H(r)\, r$ we have
\begin{equation}                                                                                                                          
\label{eq:stima}                                                                                                                          
\rho_\odot = k^2 \omega^2 (1+  2\alpha_H)                                                                             
\end{equation}                                                                                                                            
where $\omega\simeq 30\,$km/s/kpc, $k^2$ is the fraction of the halo
contribution to the circular velocity at $R_\odot$ and $\alpha_H\equiv
d{\rm log} V_H/d{\rm log} R$ is its unknown slope at the same radius.
The quantities $k^2$ and $\alpha_H$ are experimentally unknown.  One
can use different assumptions, mass modeling and data to get them, but
he/she will always find that they range from 0.3 (max disk,
\citealt{PS90}) to 0.7, and from 0.2 to 0.8 (max disk).  These are
very large variations in terms of structural properties of spirals,
but only mild ones in the determination of $\rho_\odot$, which is
dominated by the term $\omega^2$: for any reasonable value of
$\alpha_H$ and $k$, the density will be in the range
$0.25\,$GeV$<\rho_\odot<0.70\,$GeV.  An increase in precision beyond
this scale estimate would require an accurate determination of $k$ and
of the relative mass modeling, of difficult realization by means of
traditional methods as discussed in the introduction.  The present
method will be able to determine very precisely $\rho_\odot$ through
equation (\ref{eq:11bis}) if improved/new measures of the relevant
observational quantities also emerge.

We thus believe that the value given in equation (\ref{eq:result3})
reflects the present state-of-the-art knowledge of $\rho_\odot$ and of
its uncertainty, and may result in being very useful in deriving
reliable future bounds on the DM cross sections involved in direct and
indirect DM searches.

{\small \sect{Acknowledgments} GG is a postdoctoral researcher of the
  FWO-Vlaanderen (Belgium).  CFM is a postdoctoral researcher of the
  FAPESP (Brasil). FN thanks ICTP for hospitality, where this work was
  completed.}

\appendix

\section{Effect of DM halo oblateness}
\label{sec:obladi}

In this Appendix we describe the effects of the possible DM halo
oblateness on our determination of the local DM density.  We first
observe that, while extremal situations (like e.g.\ a Dark Disk) are at
odds with observations, a mild oblateness of the DM halo is not
excluded (see e.g.\ \citealt{OBrien:2010eg} for a recent review).
 
\begin{figure}[h]
\centerline{\includegraphics[width=.91\columnwidth]{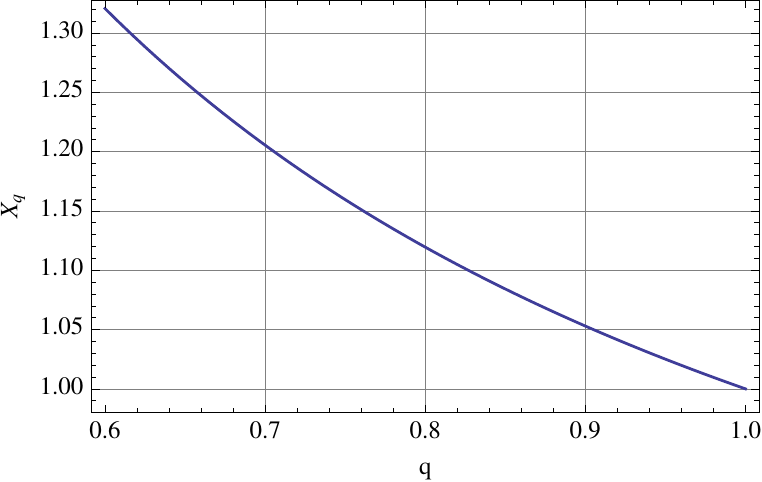}}
\caption{Effect of the DM halo oblateness $q$.}
\label{fig:obl}
\end{figure}

The effect of a halo oblateness on our method is described by the
ratio $X_q$ (included in equation~(\ref{eq:rhohalo})) between the true
local density of a halo with oblateness $q$ with the one reconstructed
via the radial derivative $R^{-2}d(RV_H^2)/dR$:
\begin{equation}
X_q=\left.\frac{\rho_{H,q}}{(4\pi G r^2)^{-1} \partial_r(rV_{H,q}^2)}\right|_{R_\odot,z=0}.
\label{eq:obl}
\end{equation}
where $V_{H,q}$ is the circular velocity produced by an oblate halo
$\rho_{H,q}$. (Detailed formulae can be found e.g.\
in~\citealt{Banerjee:2009xm}.)  Clearly $X_q$ reduces to 1 for a
spherical halo $q=1$, for which the spherical Gauss law can be used.
In figure~\ref{fig:obl} we have plotted $X_q$ as a function of $q$.
As one can see the effect is rather weak, of the order of 5\%, given a
limited oblateness of 0.9--1, near the upper end.

It is worth stressing that even a halo oblateness outside this
reference range can be straightforwardly taken into account by the
quantity $X_q$.  This is another advantage of the proposed method with
respect to the traditional one. It is in fact much easier, to
introduce a constant factor of order one in equation
(\ref{eq:rhohalo}), than to consider the effect of the halo oblateness
in fitting a global mass model, where it triggers an additional
non-uniqueness in the structural parameters.
 
\section{Effect of finite disk thickness}
\label{sec:zorro}

\begin{figure}[h]
\centerline{\includegraphics[width=.91\columnwidth]{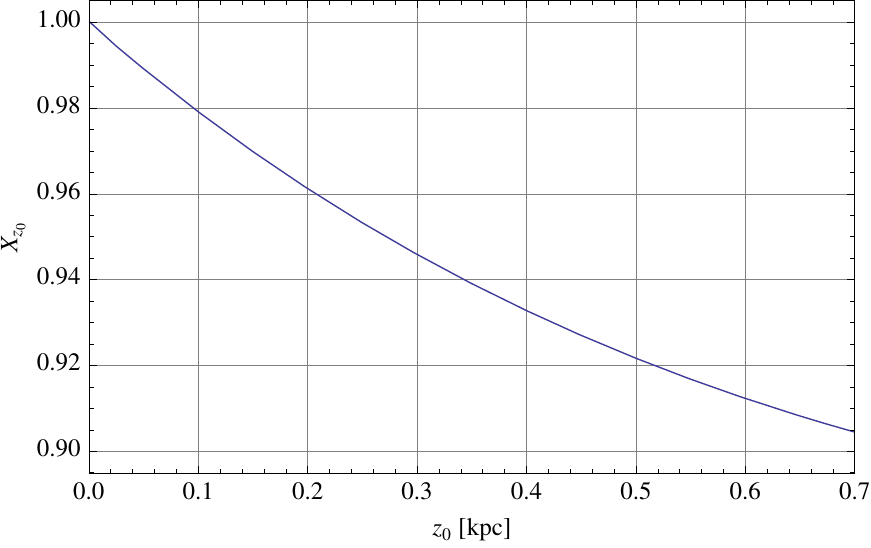}}
\caption{Effect of the disk thickness $z_0$.}
\label{fig:zorro}
\end{figure}

For the sake of completeness here we also discuss the effect of the
finite disk thickness. By neglecting it one would slightly
overestimate the acceleration produced by a given disk mass. This can
be noted from figure~\ref{fig:zorro}, where we plot a correction
factor for the `disk' part appearing on the r.h.s.\ of
equation~(\ref{eq:rhohalo}):
$$
X_{z_0}=\left.\frac{\partial_r(r^2a_{D,z_0})}{\partial_r(r^2a_{D,0})}\right|_{R_\odot,z=0}.
$$
As one can see, given that the disk thickness is measured fairly well,
$z_0=0.240$--$0.250\,$kpc~\citep{Juric:2005zr} the effect is very
weak, of the order of 5\% (with an uncertainty $<1$\%). Since this is
much less than the uncertainty on the disk mass, we neglected it in
the text, but it may be readily included in the $a_d$ term on the
r.h.s.\ of equation~(\ref{eq:rhohalo}).

\section{Issues in global approaches}
\label{app:CU}

The standard CO method reproduces by a maximum likelihood analysis a
heterogeneous set of data (among them terminal and dispersion
velocities, z-motions etc.) with an even more heterogeneous, large set
of free parameters (among them $R_\odot$, the disk and halo masses,
the anisotropy in the motions of tracers of the gravitational field
etc), after which a number of crucial assumptions are taken.  In this
way it also obtains $\rho_\odot$, as a byproduct. This strategy does
not guarantee that the global MW model obtained is physically
meaningful, even at the central best fit value of the parameters.  For
instance, the global model produces three independent and different
estimates of $V(r)$: 1) from the HI terminal velocities $V_T(r)$ (via
$V(r)= V_T(r)+ V_{\odot} \ r/R_{\odot}$), 2) from the halo stars
dispersion velocities (via the Jeans equation), and 3) from the
gravitational potential produced by the MW mass distribution. Clearly,
by physical consistency, these three estimates of $V(r)$ must agree, but
in the strategy commonly employed there is nothing forcing this to
happen.

\begin{figure}[h]
\centerline{\includegraphics[width=.97\columnwidth]{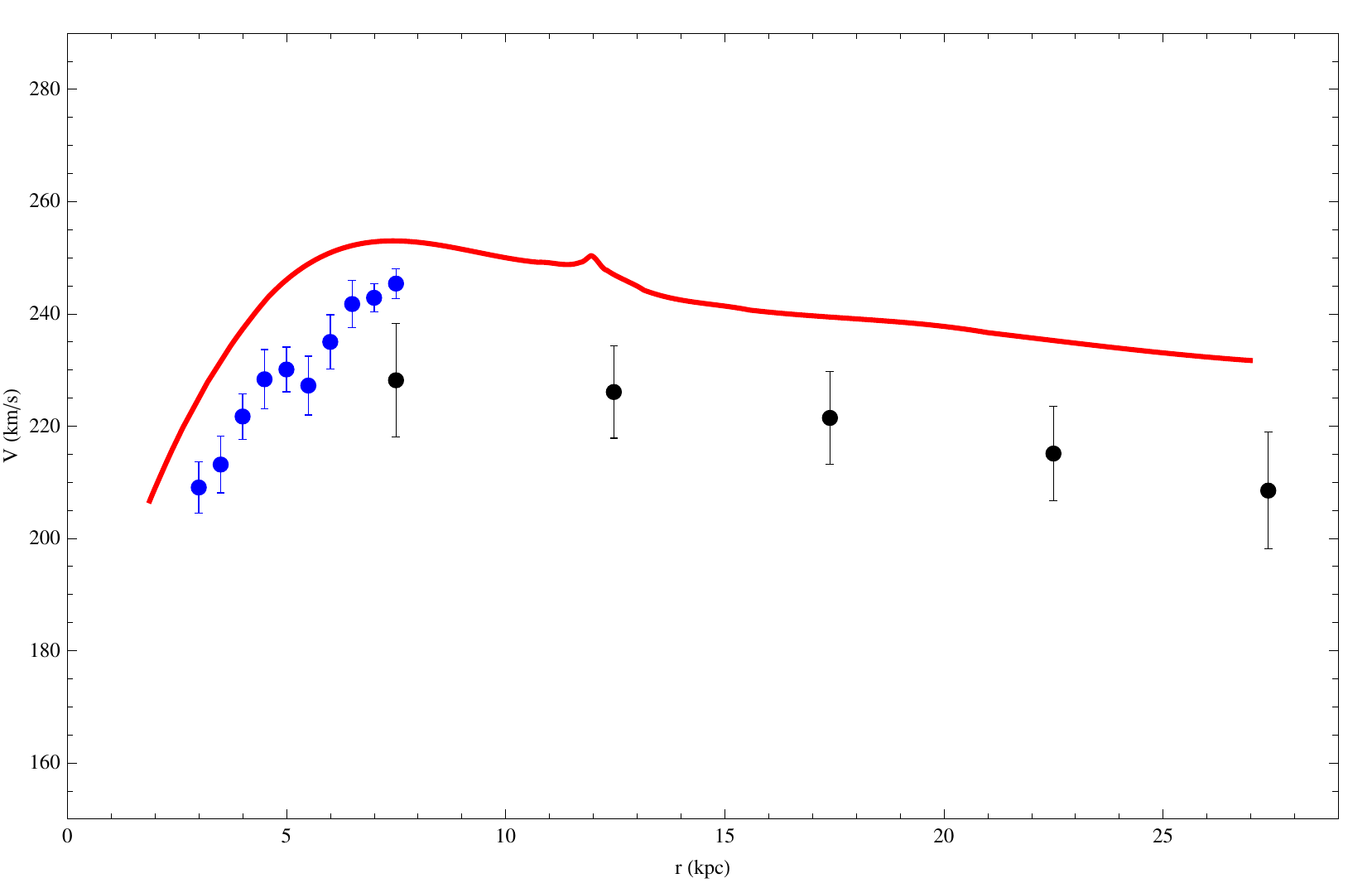}}
\vspace*{-.5ex}
\caption{Different estimates of the circular velocity of the Milky
  Way, resulting from the global mass model of~\cite{cu09}. These are
  obtained from the gravitational potential (solid line), the HI
  terminal velocities~\citep{Malhotra95} (blue points), and 
  velocity dispersions~\citep{xue08} (black points).}
\label{fig:c}
\end{figure}

In fact, let us look at the galaxy global model obtained recently by
this method by \cite{cu09}. The above 3 different estimates of $V(r)$
are obtained by using the best fit model parameters as given in their
Tables 2 and 3.  We find (figure~\ref{fig:c}) that they disagree by
10\% in amplitude and 0.3 in slope, quantities more than allowed by
the observational errors.  This implies that this method has an
intrinsic uncertainty that may lead to a biased measure of
$\rho_\odot$ and of its uncertainty.

\pagebreak[3]


\begin{thebibliography}{}

\bibitem[PDG(2008)]{pdg08}
Amsler, C., \emph{et al.},
\emph{PLB} {\bf 667} (2008) 1.  

\bibitem[Battaglia(2005)]{battaglia05}
Battaglia, G., \emph{et al.},
\emph{MNRAS} {\bf 364} (2005) 433.

\bibitem[Berezhiani et al.(2009)]{bigr}
  Berezhiani, Z., Nesti, F. , Pilo, L. and Rossi, N.,
  JHEP {\bf 0907} (2009) 083
 [arXiv:0902.0144 [hep-th]];
  Berezhiani, Z., Pilo, L. and Rossi, N.,
  arXiv:0902.0146 [astro-ph.CO].



\bibitem[Bosma et al.(1981)]{bosma81}
Bosma, A., 
\emph{AJ} {\bf 86} (1981) 1825.

\bibitem[Bovy et al.(2009)]{bovy}                                                                                             
  Bovy, J.,~Hogg,  D.W. and Rix, H.W.,                                                                              
  Astrophys.\ J.\  {\bf 704} (2009) 1704                                                                          
  [arXiv:0907.5423 [astro-ph.GA]].                                                                                

\bibitem[Brown et al.(2009)]{brown09}
  Brown, W.R., Geller, M.J., Kenyon, S.J. and Diaferio, A.,
  arXiv:0910.2242 [Unknown].

\bibitem[Caldwell, Ostriker(1981)]{caldwell81}
Caldwell, J.A.R. and Ostriker,  J.P.,
\emph{ApJ} {\bf 251} (1981) 61. 

\bibitem[Catena, Ullio(2009)]{cu09}
Catena, R. and Ullio, P., arXiv:0907.0018 [astro-ph.CO]. 

\bibitem[DeBlok(2009)]{deblok09}
de Blok, W.J.G.,
arXiv:0910.3538.

\bibitem[Donato et al.(2009)]{donato09}
Donato, F., Maurin, D., Brun, P., Delahaye, T. and Salati, P.,
\emph{PRL} {\bf 102} (2009) 071301.

\bibitem[Ellis et al.(2008)]{ellis}
Ellis, J., Olive, K.A. and Savage, C,
\emph{PRD} {\bf 77} (2008) 065026. 

\bibitem[Fall Efstathiou(1980)]{fall80}
Fall, S.M. and Efstathiou, G., 
\emph{MNRAS} {\bf 193} (1980) 189. 

\bibitem[Freudenreich(1998)]{Freudenreich}
Freudenreich, H.T., \emph{Astrophys.\ J.} 492, 495 (1998) [arXiv:astro-ph/9707340].

\bibitem[Freeman(1970)]{Freeman} 
Freeman, K.C.\ 1970, \apj, 160, 811 

\bibitem[Gentile et al.(2005)]{gentile05}
Gentile, G., Burkert, A. , Salucci, P., Klein, U. and Walter, F.,
\emph{ApJ} {\bf 634} (2005) L145. 

\bibitem[Gentile et al.(2004)]{gentile04}
Gentile, G., Salucci,  P., Klein, U., Vergani, D. and Kalberla, P.,
\emph{MNRAS}{\bf 351} (2004) 903.

\bibitem[Ghez et al.(2008)]{ghez}
  Ghez, A.M. {\it et al.},
  Astrophys.\ J.\  {\bf 689} (2008) 1044
  [arXiv:0808.2870 [astro-ph]].

\bibitem[Gillessen et al.(2008)]{Gillessen:2008qv}
  Gillessen, S., Eisenhauer, F., Trippe, S., Alexander, T., Genzel,
  R., Martins, F. and Ott, T.,
  Astrophys.\ J.\  {\bf 692} (2009) 1075
  [arXiv:0810.4674 [astro-ph]].

\bibitem[M. Juri{\'c} et al.(2008)]{juric08}
Juri{\'c}, M. , \emph{et al.},
\emph{ApJ} {\bf 673} (2008) 864.


\bibitem[Klypin et al.(2002)]{Klypin}
 Klypin, A., Zhao, H. and Somerville, R.S.,
  Astrophys.\ J.\  {\bf 573} (2002) 597.


\bibitem[Malhotra(1995)]{Malhotra95}
Malhotra, S.,
\emph{ApJ}, {\bf 448} (1995) 138.

\bibitem[McClure-Griffiths, Dichey(2007)]{McClure07}
McClure-Griffiths, N.M., and Dickey, J.M., 
\emph{ApJ} {\bf 671} (2007) 427. 


\bibitem[McMillan, Binney(2009)]{MB09}
McMillan, P.J. and Binney, J.J., arXiv:0907.4685 [astro-ph.GA]. 

\bibitem[Nakanishi, Sofue(2003)]{nakanishisofue} 
Nakanishi, H. and Sofue, Y.,
\emph{PASJ},  {\bf 55}, (2003) 191. 

\bibitem[Navarro et al.(1996)]{NFW96}
Navarro, J.F., Frenk, C.S. and White,  S.D.M.,
\emph{AJ} {\bf 462} (1996) 563.

\bibitem[O'Brien et al.(2010)]{OBrien:2010eg}
 O'Brien,  J.C., Freeman, K.C. and van der Kruit, P.C.,
  arXiv:1002.3098 [astro-ph.CO].

\bibitem[Olling, Merrifield(2001)]{Olling}
Olling, R. and  Merrifield, M., MNRAS 164 (2001) 326.

\bibitem[Persic, Salucci(1990)]{PS90}
Persic, M. and Salucci, P., 
\emph{MNRAS} {\bf 245} (1990) 577.

\bibitem[Persic, Salucci(1990)]{PS90b}
Persic, M. and Salucci, P., 
\emph{MNRAS}, {\bf 247} (1990) 349. 

\bibitem[Persic, Salucci(1996)]{PSS}
Persic, M., Salucci, P. and Stel, F.,  
\emph{MNRAS} {\bf 281} (1996) 27.

\bibitem[Picaud, Robin(2004)]{PR04}
Picaud, S. and Robin, A.C.,
\emph{A\&A} {\bf 428} (2004) 891. 

\bibitem[Reid(2009)]{reid09} 
Reid, M.J. \emph{et al.},
\emph{ApJ} {\bf 700} (2009) 137. 

\bibitem[Reyl{\'e}(2009)]{reyle09}
Reyl{\'e}, C., Marshall, D.J., Robin, A.C. and Schultheis{\'e}, M.,
\emph{A\&A} {\bf 495} (2009) 819. 

\bibitem[Robin et al.(2008)]{robin08}
Robin, A.C., Reyl\'e, C. and Marshall, D.J.,
\emph{AN} {\bf 329} (2008) 1012. 

\bibitem[Rubin et al.(1980)]{rubin80}
Rubin, V.C., Ford Jr., W.K. and Thonnard, N., 
\emph{ApJ} {\bf 238} (1980) 471.

\bibitem[Salucci et al.(2007)]{URC2}
Salucci, P., Lapi, A., Tonini, C., Gentile, G., Yegorova, I. and
Klein,  U., 
\emph{MNRAS} {\bf 378} (2007) 41.

\bibitem[Sanders, McGaugh(2002)]{mond}
Sanders, R.H. and McGaugh, S.S., 
\emph{Annual Rev.\ A\&A} {\bf 40} (2002) 263.

\bibitem[Savage et al.(2009)]{savage09} 
Savage, C., Freese, K., Gondolo, P. and Spolyar, D..
\emph{J. of Cosm.\ and Astro-Particle Phys.} {\bf 9} (2009) 36.

\bibitem[Sofue(2009)]{sofue1} 
Sofue, Y.,
\emph{PASJ},  {\bf 61}, (2009) 153. 

\bibitem[Sofue et al.(2009)]{sofue2}  
  Sofue, Y. , Honma, M. and Omodaka, T.,
 \emph{PASJ} {\bf 61} (2009) 229,
  [arXiv:0811.0859 [astro-ph]].


\bibitem[Tonini, Salucci(2004)]{Tonini} 
Tonini, C. and Salucci,  P., (2004), bdmh.conf, 89T.

\bibitem[Weber, de Boer(2009)]{weber09}
Weber, M.  and de Boer, W.,  
arXiv:0910.4272 [astro-ph.CO].

\bibitem[Xue et al.(2008)]{xue08} 
Xue, X.X. \emph{et al.}, 
\emph{ApJ} {\bf 684} (2008) 1143.

\bibitem[Juric et al.(2008)]{Juric:2005zr}
  Juric, M. {\it et al.}  [SDSS Collaboration],
  Astrophys.\ J.\  {\bf 673} (2008) 864
  [arXiv:astro-ph/0510520].
 

\bibitem[Saha et al.(2009)]{smalloblateness}
  Saha, K., Levine, E.S. , Jog, C.J. and Blitz, L.,
  Astrophys.\ J.\  {\bf 697} (2009) 2015
  [arXiv:0903.3802 [astro-ph.GA]].

\bibitem[Banerjee et al.(2009)]{Banerjee:2009xm}
  Banerjee, A. , Matthews, L.D. and Jog, C.J.,
  New Astron.\  {\bf 15} (2010) 89
  [arXiv:0906.0217 [astro-ph.CO]].


 




\end{thebibliography}
\end{document}